\documentclass[twocolumn]{my}          
\smartqed  
\usepackage{graphicx}
\usepackage{amsmath}
\usepackage{amssymb}
\usepackage{mathrsfs}
\usepackage{physics}
\usepackage{bbold}
\usepackage{dirtytalk}
\usepackage[utf8]{inputenc}
\usepackage[T1]{fontenc}
\usepackage{qcircuit}

\DeclareMathOperator\arctanh{arctanh}

\begin{document}

\title{Sustainability of Global Economy as a Quantum Circuit
}


\author{Antonino Claudio Bonan}


\institute{Antonino C. Bonan \at
              ARPAV - Servizio Meteorologico \\
              via Marconi 55, 35037 Teolo (PD), Italy \\
              \email{antoninoclaudio.bonan@arpa.veneto.it}
}

\date{}

\maketitle

\begin{abstract}
In economy, viewed as a quantum system working as a circuit, each process at the microscale is a quantum gate among agents. The global configuration of economy is addressed by optimizing the sustainability of the whole circuit. This is done in terms of geodesics, starting from some approximations. A similar yet somehow different approach is applied for the closed system of the whole and for economy as an open system. Computations may partly be explicit, especially when the reality is represented in a simplified way. The circuit can be also optimized by minimizing its complexity, with a partly similar formalism, yet generally not along the same paths.
\end{abstract}
\section{Introduction}
\label{intro}
A thinkable approach refers to economical agents and money as to quantum fields: this should not be taken as reductionism, but as the proposal of a possible point of view radically applying the formalism of theoretical physics to the foundations of economy. The configuration of such a global system depends on actual constraints, e.g. the ones dictated by (or at least compatible with) sustainability, anyway the latter is meant. The design of quantum circuits for computation is paradigmatic, as a way to plan experimental configurations in order to solve problems; in this sense, the purpose of designing the economy is linked to the optimization of the corresponding quantum circuits. Namely, after specifying actual limits and the boundaries of sustainability, constraints apply to any algorithm aimed at optimizing the whole. \\
In order to be universal at the microscale, econometry must be based on observables which are common to all agents. There is a remaining freedom to choose them: the amount of money is just an example. Macrostates then emerge in terms of different entities, such as entropy and conserved average observables of the global system of economy and nature\footnote{In Bonan (2021), economy is addressed both from a micro and a macro point of view, with the corresponding characterization in terms of observables and emergence from quantum dynamics.}. \\
Here, for the sake of simplicity, the amount of global wealth (as it is ready for any possible economic process, allocated by each snapshot of the whole economic dynamics) is chosen as the only common observable. After forcing discretization in units of an universal currency, the spectrum of its possible values is also common to all local measurements. This means that agents are identified with qudits, where \textit{d} stands for the common dimension of the corresponding Hilbert spaces. \\
Another approximating hypothesis limits the freedom to increase the number of agents within a common number \textit{n}, to be used to explicitly define the embedding quantum dynamics of the whole as a direct product of \textit{n} Hilbert spaces. \\
One could think of quantum computers as very modern tools to deal with very difficult numerical tasks in problems of economy. Here the approach is different and deals with much more global founding features. Namely, computing is addressed so that it somehow \say{faithfully} represents economic dynamics as a whole, involving all possibile agents and money. The matter is then to treat any economic configuration as a global quantum circuit; instead of its computing tasks, one focuses on the actual dynamics representing them. \\
Within the approximations above, the circuit is made of qudits and each gate linking them corresponds to a process at the microscale. As a matter of fact, all unitary operators which characterize the dynamical functioning of quantum circuits with qudits are universally made of gates with no more than two entries, as well as possible controls between qudits or changes of phase\footnote{See Luo and Wang (2014).}. Fig. 1 shows a set of primitive processes at the microscale, from which any economic configuration can be built.
\begin{figure}
	\Qcircuit @C=3.4em @R=2.5em {
& \qw & \qw && \gate{} & \qw && \ctrl{1} & \qw \\
&& && &&& \qw & \qw
}
	\caption{Primitive gates for qudits in quantum circuits, representing global configurations for the dynamics of microeconomy. From left to right, respectively: determination of the microstate for an agent, its phase change (only affecting possible interference), one agent controlling another.}
	\label{fig:1}
\end{figure}
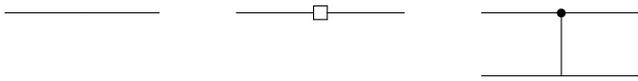
\section{General geometrical approach to sustainability of a closed system}
\label{sec:2}
Disposable resources and natural limits, drawing the boundaries of sustainability, define the constraints of global economic configurations. They can outline a thermodynamic approach, in which \textit{totally renewable resources} are meant as infinite reservoirs. Then \textit{sustainable states} are meant as the ones being equilibrated in a thermal-like fashion with such reservoirs. \textit{Sustainable processes} are paths of economic dynamics which do not pass through other types of states. Related optimization problems are well treatable, provided one properly formalizes such processes, e.g. through a generalized approach based on differential geometry. \\
Within this approach, a possible suggested \textit{measure of sustainability} is a divergence cost function which is given by the relative entropy with respect to a thermal-like equilibrium. According to the way information about microeconomy is coded in macrostates, the definition of entropy itself varies and is not necessarily of usual Von Neumann type: however, here it is just assumed to allow for deducing what is mathematically a true metric\footnote{A general formalism, to do that, can be found in Amari (2016). In the classical regime a generalized divergence function is introduced and then leads to a Fisher metric: see Rodriguez et al. (2019). In the quantum regime, one is generally led to other kinds of cost functions and therefore of metric: compare e.g. Ciaglia (2020).}. \\
Quantum circuits of global economy may be optimized, in a sense that their design may be as close as possible to sustainability, by minimizing an abstract distance defined as said above. If the system is closed as a whole, then gates are described by invertible unitary operators, while thermally equilibrated states (i.e. completely sustainable ones) evolve according to geodesics characterized by given expectation values of the global wealth. In other words: \\
a) \textit{reversibility}, as a characteristic property of processes involving closed systems (which are symmetric with respect to time reversal), is implemented by the invertibility of evolution operators; \\
b) sustainability allows for the conservation of global wealth at the macroscale, as a fixed average. \\
Out of thermal-like equilibrium, dynamics may possibly be described by geodesics as well, provided a quite different concept may be given for sustainability. In this case one should maximize the entropy with constraints on different moments (rather than the average) in the statistical distribution of the same observable, or on the expectation value of some other observables (rather than the global wealth)\footnote{See Balian (2005). Sparaciari et al. (2017) showed that, if the entropic function is additive and convex, in the limit of many copies of the system (as described e.g. by a good statistical sample) entropy and average energy together characterize any state instead of just equilibrated ones.}. \\
The whole matter is generally managed through a tricky calculation involving a suitable representation for the Lie group of gates, so that the geodesic equation is made explicit and possibly solved, with the result of a design of the global economic circuit depending on what is fixed. It mainly consists of\footnote{The formalism of Loubenets and K$\mathrm{\ddot{a}}$ding (2020) can be used in order to adapt the calculation described in Balian (2014) from qubits to qudits. It becomes quite different as one fixes averages rather than other moments.}:
\begin{itemize}
\item deducing the unitary evolution operator of the system from the Hamiltonian operator characterizing it as a quantum circuit;
\item evolving the density matrix in time through such an operator;
\item expressing the entropy as a functional of the density matrix;
\item expressing the second-order differential of entropy as a Hessian form (here comes an essential approximation, in order to neglect other kinds of dependence on the density matrix);
\item deducing the metric from the coefficients of this form;
\item formulating the geodesic equation as due to such a metric, following the analogue of a geometrical approach to general relativity.
\end{itemize}
These steps cannot be further explained for all circuits, but just for some their specific classes. In the following sections the focus is not on identifying such classes: the formalism is instead applied to a few somehow different interesting settings, possibly with proper changes.
\section{Sustainability for a closed system of one agent with two eigenstates of global wealth}
\label{sec:3}
If the economy is described by just one comprehensive agent ($n=1$) which can exactly have two fundamental states, namely survival or eclipse, then its processes are made of qubits (\textit{d=2}) and the corresponding wealth is a time dependent Hamiltonian operator of the form
\[ H (t) = h_0 (t) \sigma_0 + h_1 (t) \sigma_1 + h_2 (t) \sigma_2 + h_3 (t) \sigma_3 , \]
where a parameterization is chosen, with real coefficients $(h_{\alpha} (t))_{\alpha=0,1,2,3}$, in terms of Pauli matrices
\[ \sigma_0 = \begin{pmatrix}1 & 0\\0 & 1\end{pmatrix}, \sigma_1 = \begin{pmatrix}0 & 1\\1 & 0\end{pmatrix}, \sigma_2 = \begin{pmatrix}0 & -i\\i & 0\end{pmatrix}, \sigma_3 = \begin{pmatrix}1 & 0\\0 & -1\end{pmatrix} . \]
and the life or the end of economy are respectively represented by normalized states $\big(\begin{smallmatrix}1\\0\end{smallmatrix}\big)$ or $\big(\begin{smallmatrix}0\\1\end{smallmatrix}\big)$. \\
In this case and choosing units so that $\hslash=1$, it can be shown\footnote{See Loubenets and K$\mathrm{\ddot{a}}$ding (2020).} that the unitary operator of evolution starting from time $t_0$ is
\[
U(t,t_0) = e^{-i\int_{t_0}^{t} h_0 (\tau) \,d\tau} \left[ u(t,t_0) \sigma_0 + i \sum_{\alpha=1}^{3} v_{\alpha} (t,t_0) \sigma_{\alpha} \right]
\]
where $q = u \oplus v_1 \oplus v_2 \oplus v_3$ is a complex $4$-vector function satisfying the differential equation
\[ \frac{dq(t,t_0)}{dt} = \begin{pmatrix}0 & h_1 (t) & h_2 (t) & h_3 (t)\\-h_1 (t) & 0 & -h_3 (t) & h_2 (t)\\-h_2 (t) & h_3 (t) & 0 & -h_1 (t)\\-h_3 (t) & -h_2 (t) & h_1 (t) & 0\end{pmatrix} q(t,t_0) \]
with the initial conditions $q_{\alpha} (t_0 , t_0) = \delta_{\alpha}^0$ for $\alpha=0,1,2,3$. \\
Furthermore, if the wealth is an observable with fixed eigenvalues $E_a > E_d$ respectively associated to $\big(\begin{smallmatrix}1\\0\end{smallmatrix}\big)$ and $\big(\begin{smallmatrix}0\\1\end{smallmatrix}\big)$, then one has $\forall t$
\[ h_0 (t) = E_m , h_1 (t) = h_2 (t) = 0 , h_3 (t) = \Delta , \]
with $E_m=$\(\frac{E_a+E_d}{2}\) and $\Delta=$\(\frac{E_a-E_d}{2}\), so that the differential equation above is solved by
\[ q (t,t_0) = \begin{pmatrix}\cos{\left[(t-t_0)\Delta\right]} \\ 0 \\ 0 \\ -\sin{\left[(t-t_0)\Delta\right]}\end{pmatrix} \]
and the unitary operator of evolution is given by
\[
\begin{split}
U(t,t_0) & = \exp\left[-i(t-t_0)E_m\right] \cdot \\
 & \quad \cdot \left\{ \sigma_0 \cos{\left[(t-t_0)\Delta\right]} + \right. \\
  & \left. \qquad + i \sigma_3 \sin{ \left[ (t-t_0)\Delta \right] } \right\} .
\end{split}
\]
Density matrices represent statistical distributions mixing microstates and can be written as follows in terms of real coefficients $r^j(t)$ for $j=1,2,3$ (where upper indexes are not exponents):
\[ \rho (t) = \frac{1}{2}[\sigma_0 + r^1 (t) \sigma_1 + r^2 (t) \sigma_2 + r^3 (t) \sigma_3] . \]
They evolve according to
\[ \rho (t) = U (t,0) \rho(0) U^+ (t,0) \]
so that
\[ \begin{pmatrix} r^1(t)\\ r^2(t) \\ r^3(t) \\ \end{pmatrix} = \begin{pmatrix} r^1(0)\cos(2t\Delta)+r^2(0)\sin(2t\Delta)\\ r^2(0)\\r^3(0)\\ \end{pmatrix} \]
with the choice $t_0=0$. At this stage, it is worth noting that the dynamics represented by $\Vec{r}(t)$ exhibits a period $\pi/\Delta$ and is modulated along $\sigma_1$ only, depending upon the components of $\Vec{r}(0)$ along $\sigma_{1,2}$. \\
The eigenvalues of $\rho(t)$ are $\xi_{1,2}\{\rho(t)\}=(1\pm\| \Vec{r}(t) \|)/2$ and are useful for the computation of the corresponding von Neumann's entropy $S\{\rho(t)\}$ given by
\[ S[\Vec{r}(t)]= -\xi_1\{\rho(t)\}\log\xi_1\{\rho(t)\}-\xi_2\{\rho(t)\}\log\xi_2\{\rho(t)\} . \]
From this $S$, the metric of sustainability is found for any given time\footnote{Here the calculation by Balian (2014) is followed.} in covariant indexes as the $3\times3$ matrix of coefficients in the Hessian form $-d^2S$:
\[
\begin{split}
g_{jk} & = \left(-\frac{\partial^2S[\Vec{r}]}{\partial r^j \partial r^k}\right) \\
 & = \frac{r^jr^k}{\|\Vec{r}\|^2}\left(\frac{1}{1-\|\Vec{r}\|^2}-\frac{\arctanh{\|\Vec{r}\|}}{\|\Vec{r}\|}\right)+\delta_j^k\frac{\arctanh{\|\Vec{r}\|}}{\|\Vec{r}\|} ,
\end{split}\]
while in contravariant indexes
\[
\begin{split}
g^{jk} & = \left(-\frac{\partial^2F[\Vec{\chi}]}{\partial \chi_j \partial \chi_k}\right) \\
 & = (1-\|\Vec{r}\|^2)\frac{r^jr^k}{\|\Vec{r}\|^2}+\frac{\|\Vec{r}\|}{\arctanh{\|\Vec{r}\|}}\left(\delta_j^k-\frac{r^jr^k}{\|\Vec{r}\|^2}\right),
\end{split}\]
where
\[ \chi_j =-\frac{\partial S[\Vec{r}]}{\partial r^j} = \frac{r^j\arctanh{\|\Vec{r}\|}}{\|\Vec{r}\|} \]
and
\[ F[\Vec{\chi}]=S[\Vec{r}]+\sum_{j=1}^3 r^j \chi_j\]
is the corresponding Legendre transform of $S[\Vec{r}]$.
The Christoffel symbol is
\[
\begin{split}
\Gamma_{jkl} & = \left(-\frac{1}{2}\frac{\partial^3S[\Vec{r}]}{\partial r^j \partial r^k \partial r^l}\right) \\
 & = \frac{r^j\delta_k^l+r^k\delta_j^l+r^l\delta_j^k}{2\|\Vec{r}\|^2}\left(\frac{1}{1-\|\Vec{r}\|^2}-\frac{\arctanh{\|\Vec{r}\|}}{\|\Vec{r}\|}\right) + \\
 & \quad + \frac{r^jr^kr^l}{2\|\Vec{r}\|}\frac{d}{d\|\Vec{r}\|}\left[\frac{1}{\|\Vec{r}\|^2}\left(\frac{1}{1-\|\Vec{r}\|^2}-\frac{\arctanh{\|\Vec{r}\|}}{\|\Vec{r}\|}\right)\right]
\end{split}\]
and then brings to a geodesic equation (for $j=1,2,3$)
\[ \frac{d^2r^j}{dt^2} + \sum_{k,l,m=1}^3 g^{jk} \Gamma_{klm} \frac{dr^l}{dt} \frac{dr^m}{dt} = 0. \]
Each component of the l.h.s. in this equation is the component of a vector, whose magnitude is thus an estimate of unsustainability $A[r^1(0),r^2(0)]$ at any time and with a given half range $\Delta$ of wealth. Fig. 2 shows some resulting sustainable patterns, as obtained through numerical computation for a single period during $\pi/\Delta$. \textit{Small initial mixings and small ranges of global wealth appear as the ideal conditions for sustainability}: the former work beyond periodicity, the latter avoid the collapse of chances.
\begin{figure*}
    \centering
	\includegraphics[width=0.2\textwidth]{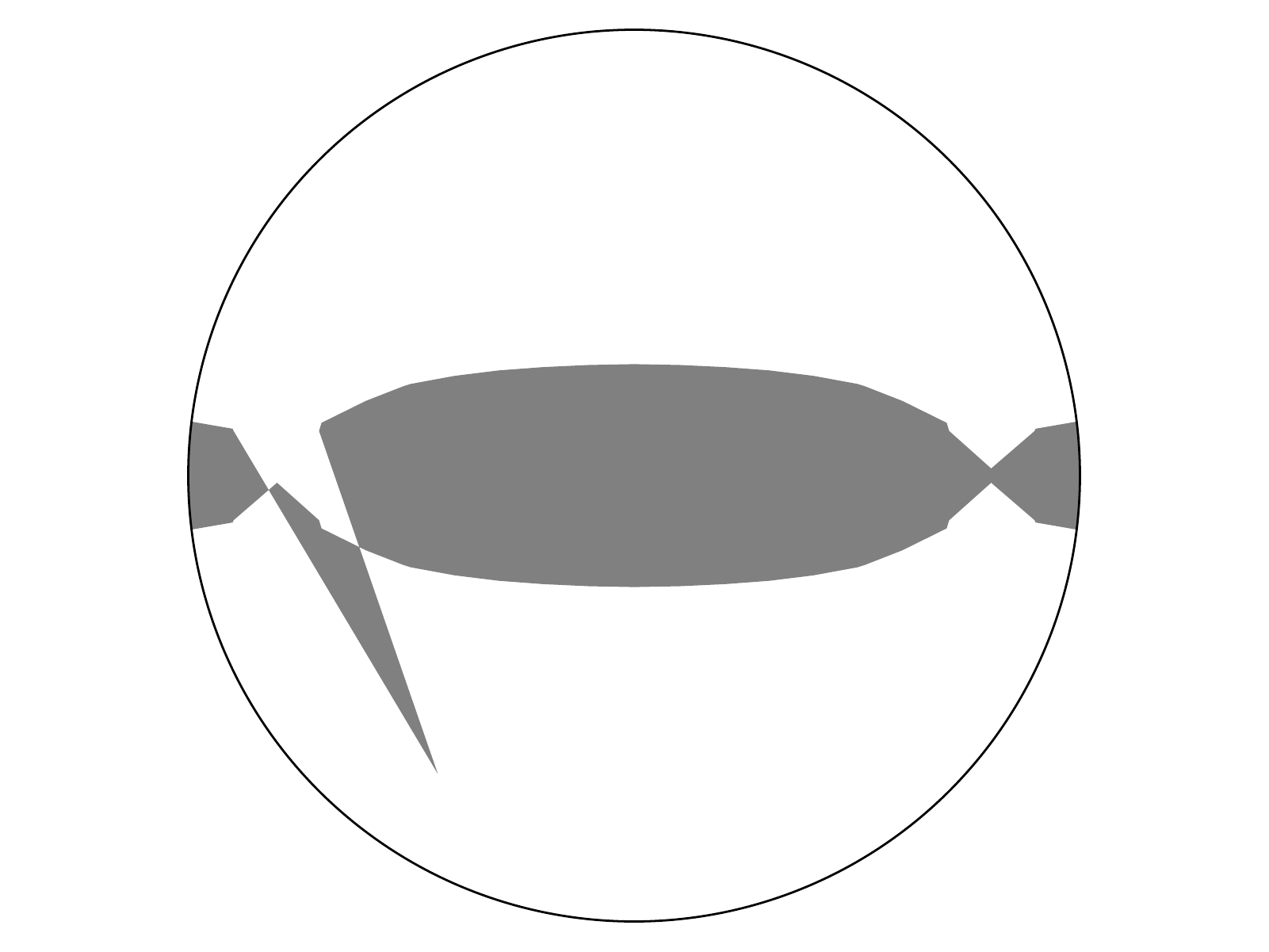}\hfill
	\includegraphics[width=0.2\textwidth]{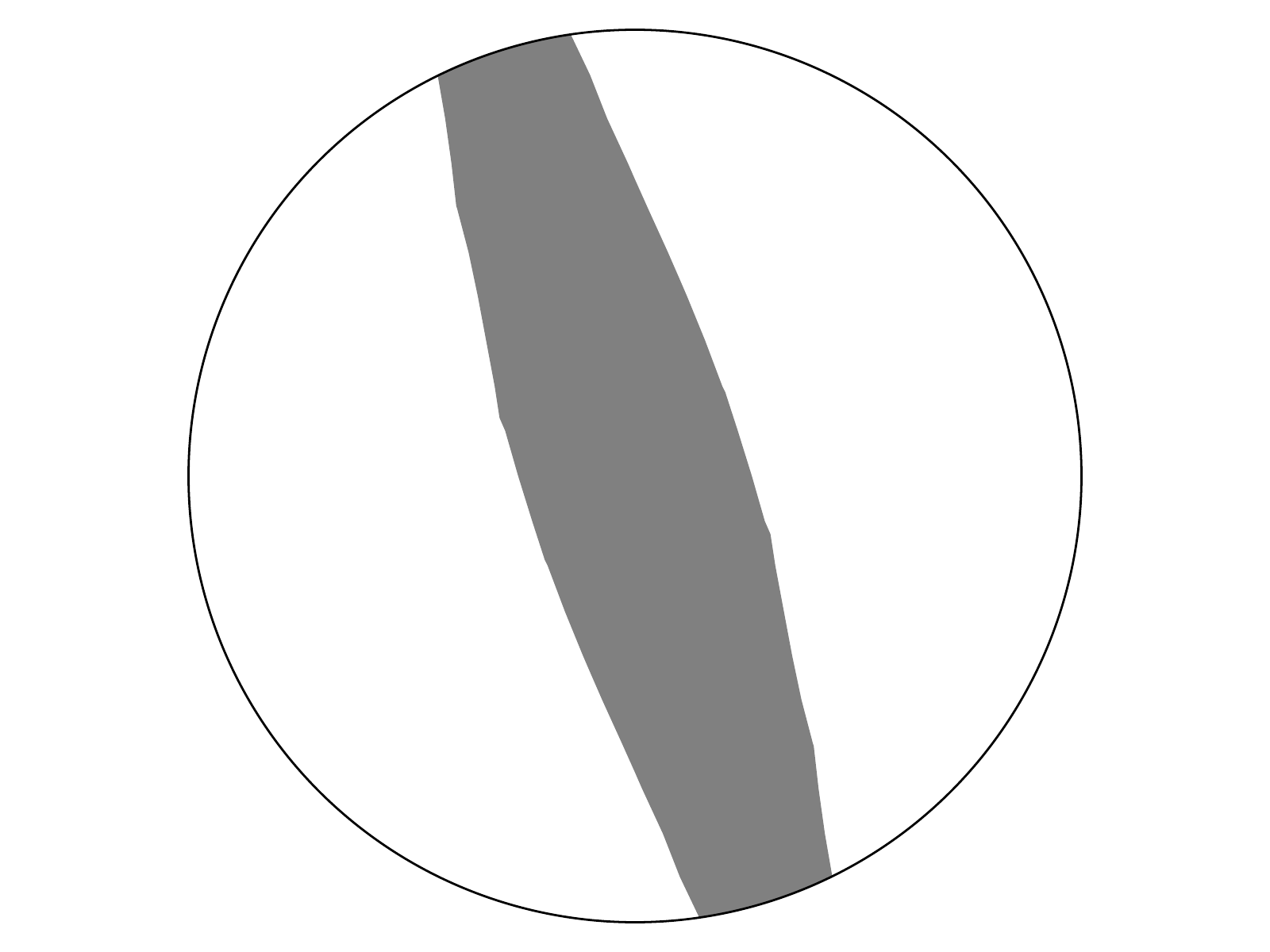}\hfill
	\includegraphics[width=0.2\textwidth]{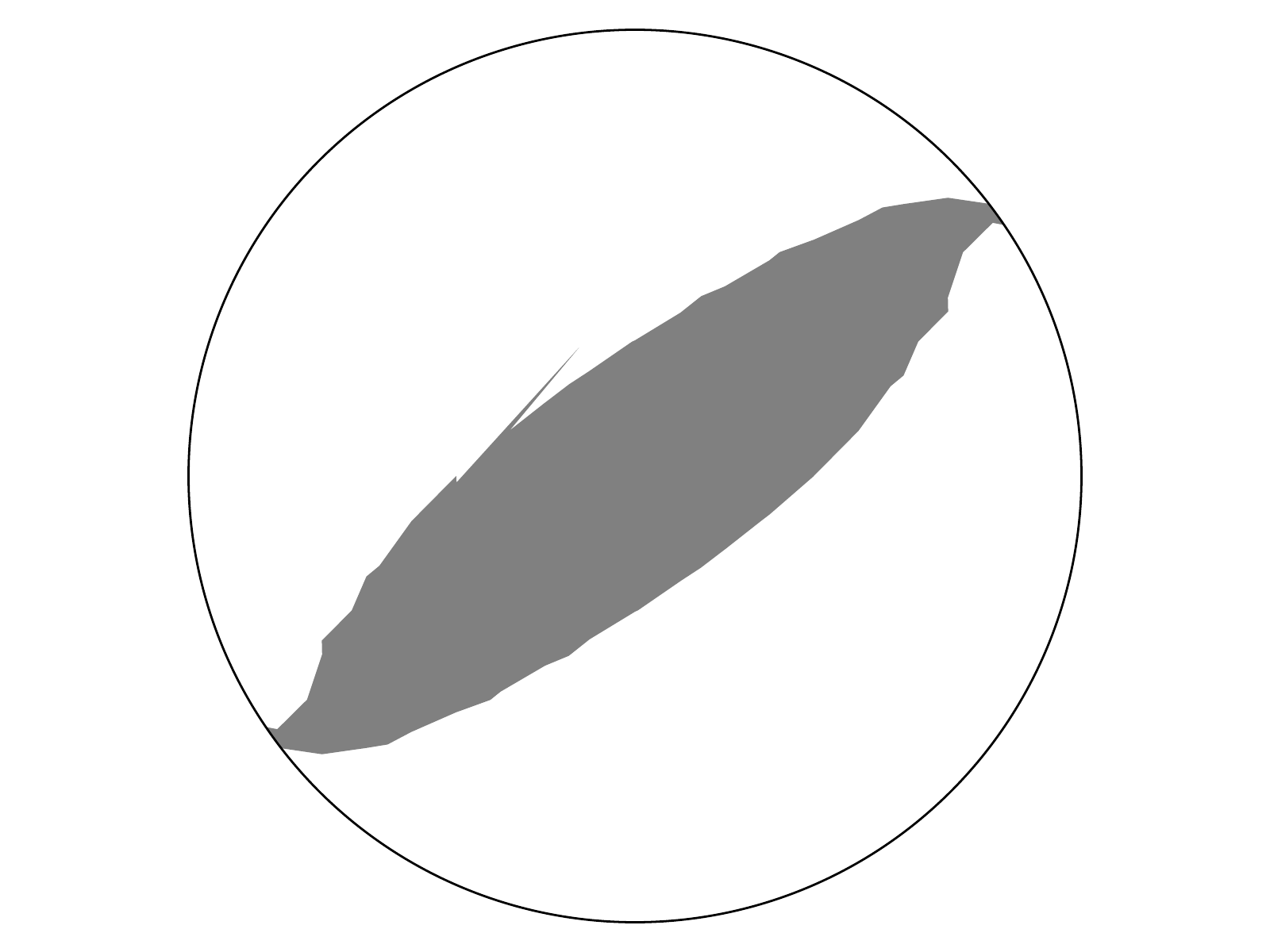}\hfill
	\includegraphics[width=0.2\textwidth]{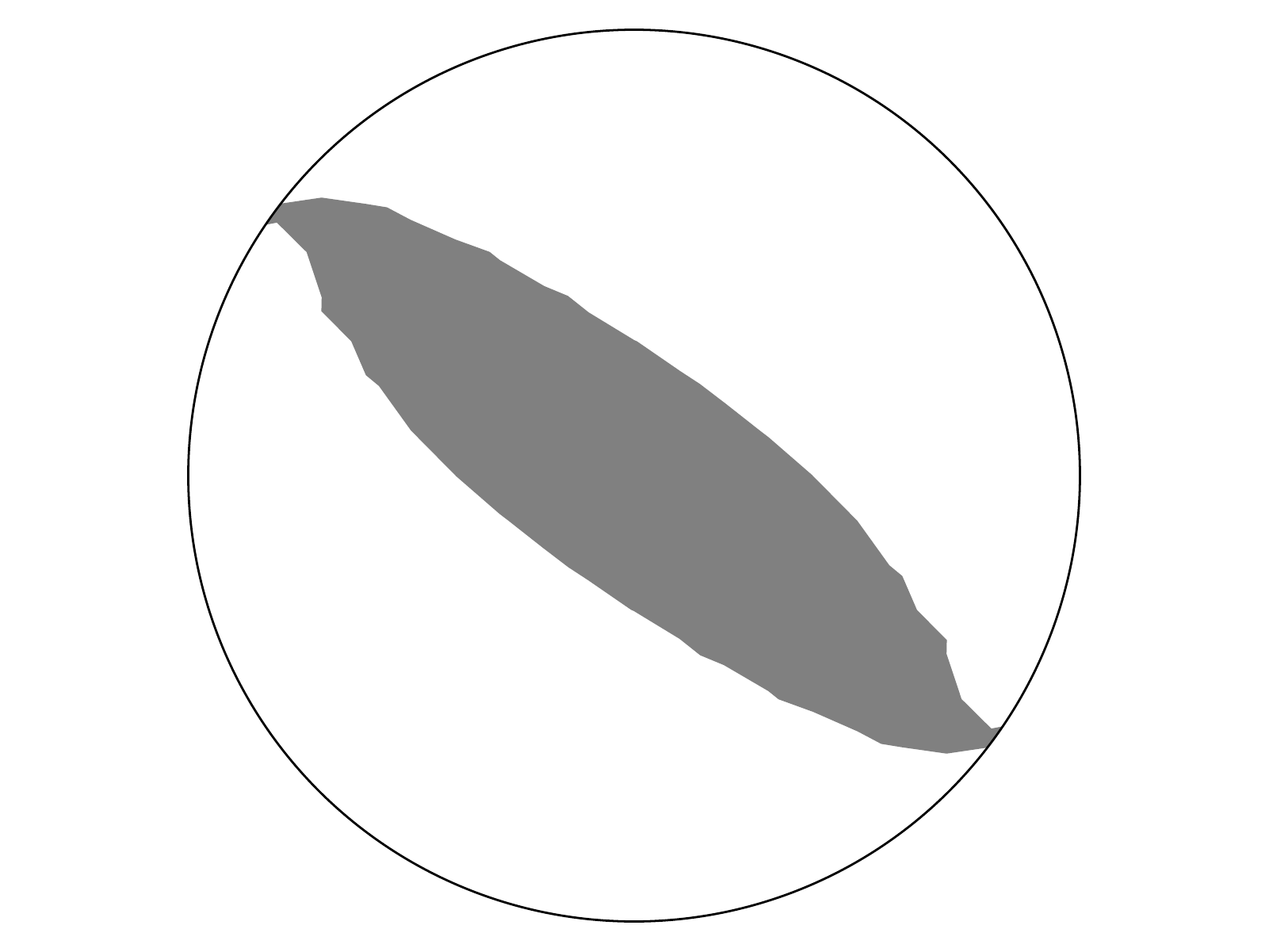}\hfill
	\includegraphics[width=0.2\textwidth]{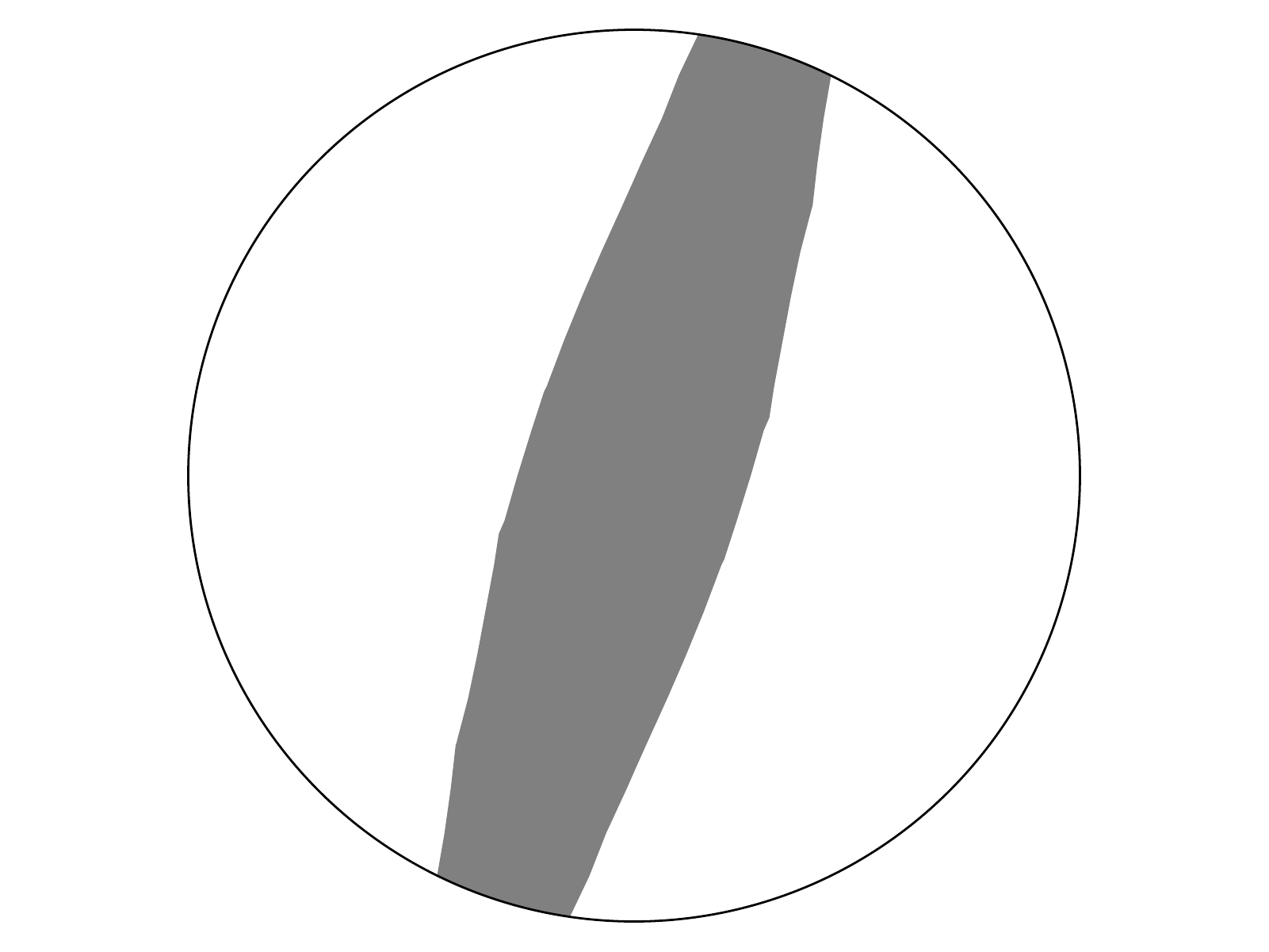}\hfill
	\includegraphics[width=0.75\textwidth]{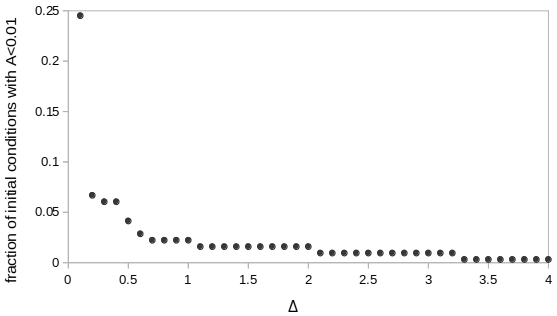}\hfill
	\caption{Some numerically estimated patterns of sustainability in a period, according to various initial mixing distributions $(r^1(0),r^2(0))$ of economical microstates and half ranges $\Delta$, for one agent and global wealth with only two eigenstates. Upper plots: grey shading for values of unsustainability $A<0.01$, enclosed in circles $\|\Vec{r}(0)\|<1$ of acceptable initial mixings, without axis labels for clarity; from left to right $t=0, 0.2(\pi/\Delta), 0.4(\pi/\Delta), 0.6(\pi/\Delta), 0.8(\pi/\Delta)$ where $\Delta=0.1$. Bottom plot quantifies the collapse of these patterns with increasing $\Delta$, as they are evaluated at half period.}
	\label{fig:2}
\end{figure*}
\section{Outlook of sustainability from historically estimated initial states}
\label{sec:4}
For the case in the previous section and within a tomographic approach starting from a statistical sample of global economic estimates, the initial density matrix can be reconstructed. There are several ways to do that\footnote{Performances of methods for a qubit in Bloch representation are well explained and compared by Schmied (2014).}, yet here just a method is suggested. \\
The probability $(1\pm r^3(0))/2$ (of initial economy being respectively alive or not) can be sampled e.g. by what happened to independent civilizations of that time (that is obviously harder in case of strong interdependence). In what is considered the same quantum state, each single separate economy is alive or dead. Their respective counts $N_{a,d}^{(0)}(0)$ lead to the expectation value $(N_a^{(0)}(0)-N_d^{(0)}(0))/N^{(3)}(0)$ of the probability to survive, where $N^{(3)}(0)=N_a^{(0)}(0)+N_d^{(0)}(0)$. Here it is assumed that such expectation value is a good unbiased estimate of its quantum counterpart $\Tr [\rho(0)\sigma_3]$.\\
Samples of the same type can be thought for other projective measurements, e.g. for responses to natural forcing, provided they act as collapses of the quantum state into subspaces with a fixed axis. This leads to make the assumption as above for the estimation of all three $\Tr [\rho(0)\sigma_j]$, $j=1,2,3$, not only for $j=3$. In summary, a simple tomographic method to reconstruct the initial quantum state from the three projective measurements along the axes of Bloch space is given by the formula of direct inversion
\[ r^j(0)=\frac{N_a^{(j)}(0)-N_d^{(j)}(0)}{N^{(j)}(0)},\quad j=1,2,3, \]
as if two independent responses of economy to forcing ($j=1,2$) were formalized just like its survival (whose two possible responses are $a,d$). The selection of the relevant forcing is then a crucial point, together with the sampling in separate economies, for a full tomography of the global initial state. This is why such an estimate is hard in the real world, where many natural and economical components of the system interact with each other. \\
Anyway, once $\Vec{r}$ is possibly reconstructed for initial time, the subsequent trajectory of sustainability for the economy of a global civilization can be computed as explained in the previous section.
\section{Possible geodesics for the sustainability of an open system}
\label{sec:5}
From a different point of view, economy is an open system: it is what can be controlled, while nature interacts with it and is what controls it. In this case, the optimization procedure has to be changed, involving another cost function whose time integral is to be minimized along the geodesic paths of dynamics. \\
If the global wealth is linearly forced by a set $X_j , j = 1, 2, \dots \alpha$ of observables according to
\[ H(t) = \sum_{j=1}^{\alpha} \lambda^j(t) X_j , \]
then\footnote{See Abiuso et al. (2020).} up to higher order corrections with respect to quasi-staticity (i.e. nearly constant $\Vec{\lambda}$) the economy loses its wealth by unsustainability, against an equilibrium parameterized by the analogue $\beta$ of the $1/k_BT$ of thermodynamics, at a rate
\[ \Dot{\omega} (t) = \sum_{j,k=1}^{\alpha} \frac{d\lambda^j(t)}{dt} \widetilde{g}_{jk}[\Vec{\lambda}(t)] \frac{d\lambda^k(t)}{dt} \]
where
\[ \widetilde{g}_{jk}[\Vec{\lambda}(t)]= \beta \left[\frac{\partial}{\partial \lambda^j(t)} \frac{\partial}{\partial \lambda^k(t)} \Dot{\omega}(t) \right]_{\frac{\partial \Vec{\lambda}(t)}{\partial t}=\Vec{0}} \]
is a well defined non negative metric whose inverse is $\widetilde{g}^{jk}[\Vec{\lambda}(t)]$. If one wants to minimize this disequilibrium, then Riemannian geometry leads to a geodesic equation
\[ \frac{d^2\lambda^j(t)}{dt^2} + \sum_{k,l=1}^{\alpha} \widetilde{\Gamma}^j_{kl}[\Vec{\lambda}(t)] \frac{d\lambda^k(t)}{dt} \frac{d\lambda^l(t)}{dt} = 0 ,\]
where the Levi-Civita connection is
\[
\begin{split} \widetilde{\Gamma}^j_{kl}]\Vec{\lambda}(t)] & = \frac{1}{2} \sum_{m=1}^{\alpha} \widetilde{g}^{jm}[\Vec{\lambda}(t)] \cdot \\  & \qquad \cdot \left[\frac{\partial\widetilde{g}_{km}}{\partial \lambda^l}[\Vec{\lambda}(t)]+\frac{\partial\widetilde{g}_{lm}}{\partial \lambda^k}[\Vec{\lambda}(t)]-\frac{\partial\widetilde{g}_{kl}}{\partial \lambda^m}[\Vec{\lambda}(t)]\right] .
\end{split} \]
Some approximations could allow for an explicit application of the above Riemannian formalism even for an open system being out of equilibrium. Yet in general, due to quantum deviations from the second law of thermodynamics, total entropy production may be negative\footnote{See Funo et al. (2018).} (that is, the whole system may organize itself at the micro scale, through or without economy) and so may hinder geometrical formulations based upon a non negative metric. This is why macro is the preferred point of view in which sustainability is geometrically applied to economy as an open system. \\
A comparison with the previous sections shows that, in this case, the role of entropy and density matrices is respectively played by the rate of global wealth loss $\Dot{\omega}(t)$ and the time-dependent forcing coefficients $\Vec{\lambda}(t)$.
\section{Trajectories of sustainable simplicity}
\label{sec:6}
Another kind of geodesics can be considered, different from the one of sustainability: one can think of \textit{simplicity} (opposite to \textit{complexity}), meant as the property of circuits to minimize the number of primitive gates they are made of. A global economic system is simple to the extent that it is made of a small number of micro processes chosen in Fig.1. In the case of qubits (i.e. again for a closed system with a two-valued global wealth, but now with any number of agents), as shown below, one can get an explicit form for the corresponding geodesic equation\footnote{See Nielsen et al. (2006), even if a slightly different definition is used there for penalty. The Euler-Arnold equation of geodesics, as formulated in Arnold and Khesin (1998), is used there in order to derive main results.}. \\
One has to notice first that for $n>2$ the Hamiltonian can be written as $H=P[H]+Q[H]$, where the $P,Q$ superoperators are respectively due to gates with no more than two entries and to the other types of gates. Next, the time integrals of a cost functional $\mathcal{F}[H]=\langle H, H \rangle$ are the abstract lengths to be minimized for paths of simplicity, where
\[ 
\begin{split}
\langle J_{(1)},J_{(2)} \rangle & = \sum_{j,k=1}^{4^n-1}\widehat{g}^{jk}(q)J_{(1)j}J_{(2)k} = \\
 & =\frac{1}{2^n}\Tr{J_{(1)}P(J_{(2)})}+q\Tr{J_{(1)}Q(J_{(2)})} 
\end{split}\]
is a scalar product for any $J_{(1,2)}$ in the Lie algebra $su(2^n)$ and defines a Riemann metric $\widehat{g}$ depending on the penalty factor $q$ (i.e. the weight of gates with more than two entries in the cost). Here the sum is on the $4^n-1$ tensor products of $n$ Pauli matrices $\sigma_{0,1,2,3}$, excluding the trivial products of all $\sigma_0$; these tensor products give a basis for $su(2^n)$, so that one can define coordinates $x^{k}$ for $k=1,2,\dots ,4^n-1$. \\
Then, the following geodesic equation is obtained:
\[ \frac{d^2x^j}{dt^2} + \sum_{k,l=1}^{4^n-1} \widehat{\Gamma}^j_{kl}(q) \frac{dx^k}{dt} \frac{dx^l}{dt} = 0 ,\]
where the metric is
\[ \widehat{g}^{jk}(q) = \widehat{g}_{jk}(q) = \left\{
    \begin{array}{ll}
        0 & \mbox{if }j \neq k ,\\
        1 & \mbox{if }j=k \wedge \mbox{one or two entries} ,\\
        q & \mbox{if }j=k \wedge \mbox{three or more entries,}
    \end{array}
\right. \]
and the Levi-Civita connection is
\[ \widehat{\Gamma}^j_{kl}(q) = \frac{1}{2} \sum_{m=1}^{4^n-1} \widehat{g}^{jm}(q) \left[\frac{\partial\widehat{g}_{km}(q)}{\partial x^l}+\frac{\partial\widehat{g}_{lm}(q)}{\partial x^k}-\frac{\partial\widehat{g}_{kl}(q)}{\partial x^m}\right] .\]
For $n<3$ simplicity is ensured; this holds for all trajectories of global economy as a whole ($n=1$) and for an economy that is separated in $n=2$ independent parts. If such parts are $n>2$, then the problem of complexity arises and can be handled as above. \\
In order to \textit{minimize the complexity while saving the sustainability} of trajectories, in general, one should differentiate the geodesic equation of simplicity with respect to $q$ and then veer such a variation to satisfy the geodesic equation of sustainability. For the case of $n$ qubits (i.e. $n$ agents in an economy having just two eigenstates of global wealth), whose operators $V=e^{-ix} \in SU(n)$ are parameterised by $x(q,t) \in su(n)$, this leads to describe the corresponding variation of the circuit by the differential equation\footnote{The full computation is in Brandt (2010).}
\[ 
\begin{split}
\frac{dK}{dt} & = i(q-1)F\left([Q[H],P[K]]-[P[H],Q[K]]\right) - \\
 & \quad -F^2i[P[H],Q[H]] ,
\end{split}\]
where $[\cdot,\cdot]$ is the commutator and
\[ F=P+\frac{1}{q}Q ,\quad J(t)=\left[\frac{\partial x(q,t)}{\partial q}\right]_{q=0} ,\quad K=V\frac{dJ}{dt}V^+ .\]
This formally \say{solves} the problem of maintaining trajectories for the simplest extraction of wealth from renewable natural resources in terms of the minimization of complexity for a sustainable open system. If the resources are not renewable instead, then the problem is to be addressed by regarding the system as closed. Sustainability is a fixed property of nature for the former case, of the whole system and then the whole process for the latter case. The seemingly subtle difference is respectively between \say {exploitation of sustainable resources} and \say{sustainable exploitation of resources}. The two trajectories minimizing complexity are generally different in the space of possible configurations of the circuit.
\section{Further interpretations and final remarks}
\label{sec:7}
Here a set of three agents is first considered, again with a two-valued spectrum for global wealth. The latter operator is reconstructed\footnote{A reconstruction algorithm could be the one described in Bairey et al. (2019).} as the generator of the historical evolution of micro economical processes, i.e. of gates. It is the sum of a $Q[H]$ term for gates involving all agents at once (generating the collective evolution of the whole system) and a $P[H]$ term for others (generating the micro economical evolution of single agents and their bilateral relationships). \\
In absence of regulatory efforts collectively involving all agents, $Q=0$ leads to constant $K$ and $J$, so that no trajectory of processes is actually affected by the search for minimum complexity. On the contrary, such efforts may allow for dynamical changes aiming to possible optimal simplification. This holds with or without saving sustainability, which is an additional goal. \\
The same happens when the entire micro economy is purely collective, i.e. $P=0$: single-agent or bilateral processes are also necessary for goals of optimal simplification. \\
In summary, for $n=3$ and $d=2$, \textit{both laissez-faire and purely collective economies hinder any possible effort of simplification}. \\
If $n>3$, then things may go in a partly different way. E.g., appropriate multilateral but not fully collective processes (represented in a $P$ which is not zero even excluding single and bilateral components) may be enough for optimal simplification. \\
Sustainability of the global economy, seen as a whole without externalization of entropy, is described by a single-agent closed system. Once history has let its initial state be reconstructed, the subsequent sustainable $(E_M,\Delta)$ can be estimated along a geodesic trajectory. \\
Sustainability for multiple agents should be treated otherwise, maybe in a similar though trickier way, but is not explicitly shown in this paper. \\
In order to separate purely economical dynamics, one must think of economy as an open system, so allowing for the externalization of entropy to nature. Then sustainability can be again treated in terms of geodesic trajectories, within a quasi static description of a macro economy being linearly forced by some natural observables. Micro economy should be filtered away from these dynamics, as quantum deviations from the second law of thermodynamics would lead to possible negative entropy production. \\
As a matter of fact, natural forcings are not generally linear, while history has apparently brought mankind to heavy externalization of entropy. At the same time, significant self-organization (negative entropy production) is realized and may be interpreted as a quantum effect, which may be treated just within the formalism of circuits for quantum computation. So \textit{the optimal design of a sustainable economy as a quantum circuit is a guidance for the optimization of its self-organization}. \\
Anyway, at the present state of knowledge, the above analysis should be applied as a purely qualitative approach.

\begin{thebibliography}{}
%
%
\bibitem{RefJ}
Abiuso P, Miller HJD, Perarnau-Llobet M and Scandi M (2020), Geometric Optimization of Quantum Thermodynamic Processes, Entropy 22, 1076
\bibitem{RefB}
Amari S-i (2016), Information Geometry and Its Applications, 387 pages, Springer KK, Tokyo, Japan
\bibitem{RefB}
Arnold VI and Khesin BA (1998), Topological Methods in Hydrodynamics, Applied Mathematical Sciences, vol. 125, 376 pages, Springer-Verlag, New York, USA
\bibitem{RefJ}
Balian R (2005), Information in Statistical Physics, Studies in History and Phylosophy of Modern Physics 36, 323-353
\bibitem{RefJ}
Balian R (2014), The Entropy-Based Quantum Metric, Entropy 16, 3878-3888
\bibitem{RefJ}
Bairey E, Arad I and Lindner NH (2019), Learning a Local Hamiltonian from Local Measurements, Physical Review Letters 122, 020504
\bibitem{RefJ}
Bonan AC (2021), Some Hints about Measures on Economy as an Open Quantum System, to appear
\bibitem{RefB}
Brandt HE (2010), Quantum Computational Geodesics, Army Research Laboratory technical report 5055, 38 pages, Adelphi, USA
\bibitem{RefJ}
Ciaglia FM (2020), Quantum States, Groups and Monotone Metric Tensors, The European Physical Journal Plus 135, 1-16
\bibitem{RefB}
Funo K, Ueda M and Sagawa T (2018), Quantum Fluctuation Theorems. In: Binder F, Correa L, Gogolin C, Anders J and Adesso G (eds), Thermodynamics in the Quantum Regime. Fundamental Theories of Physics, vol. 195. Springer, Cham, Switzerland
\bibitem{RefJ}
Loubenets ER and K$\mathrm{\ddot{a}}$ding C (2020), Specifying the Unitary Evolution of a Qudit for a General Nonstationary Hamiltonian via the Generalized Gell-Mann Representation, Entropy 22, 521
\bibitem{RefJ}
Luo M, Chen X, Yang Y and Wang X (2014), Geometry of Quantum Computation with Qudits, Scientific Reports 4, 4044
\bibitem{RefJ}
Luo M and Wang X (2014), Universal Quantum Computation with Qudits, Science China Physics, Mechanics and Astronomy 57, 1712-1717
\bibitem{RefJ}
Nielsen MA, Dowling MR, Gu M and Doherty AC (2006), Quantum Computation as Geometry, Science 311, 1133-1135
\bibitem{RefJ}
Rodriguez M, Romaniega A and Tempesta P (2019), A New Class of Entropic Information Measures, Formal Group Theory and Information Geometry, Proceedings of the Royal Society A 475, 20180633
\bibitem{RefJ}
Schmied R (2014), Quantum State Tomography of a Single Qubit: Comparison of Methods, Journal of Modern Optics 63, 1744-1758
\bibitem{RefJ}
Sparaciari C, Oppenheim J and Fritz T (2017), A Resource Theory for Work and Heat, Physical Review A 96, 052112
\end{thebibliography}


\end{document}